\documentclass[prl,reprint,amsmath,amssymb,floatfix,citeautoscript,superscriptaddress]{revtex4-1}
\usepackage{cancel}
\usepackage{amsmath}
\usepackage{graphicx}
\usepackage{amssymb}
\usepackage{amsthm}
\usepackage{bm}
\usepackage{dcolumn}
\usepackage{physics}
\usepackage{longtable}

\usepackage{txfonts}

\usepackage{color}
\usepackage[usenames,dvipsnames]{xcolor}
\definecolor{myblue}{rgb}{0,0,1}
\usepackage[breaklinks=true,colorlinks=true,linkcolor=myblue,urlcolor=myblue,citecolor=myblue]{hyperref}

\let\vr\undefined
\newcommand{\vr}{{\bm{r}}}

\begin{document}

\title{
Simulations of Trions and Biexcitons in Layered Hybrid Organic-Inorganic Lead Halide Perovskites
}

\author{Yeongsu Cho}
\author{Samuel M. Greene}
\affiliation{Department of Chemistry, 
Columbia University, New York, New York 10027, USA}
\author{Timothy C. Berkelbach}
\affiliation{Department of Chemistry, 
Columbia University, New York, New York 10027, USA}
\affiliation{Center for Computational Quantum Physics, Flatiron Institute, New York, New York 10010, USA}
\email{tim.berkelbach@gmail.com}

\begin{abstract}
Behaving like atomically-precise two-dimensional quantum wells with
non-negligible dielectric contrast, the layered HOIPs have strong electronic
interactions leading to tightly bound excitons with binding energies on the
order of 500~meV.  These strong interactions suggest the possibility of larger
excitonic complexes like trions and biexcitons, which are hard to study
numerically due to the complexity of the layered HOIPs.  Here, we propose and
parameterize a model Hamiltonian for excitonic complexes in layered HOIPs and we
study the correlated eigenfunctions of trions and biexcitons using a combination
of diffusion Monte Carlo and very large variational calculations with explicitly
correlated Gaussian basis functions.  Binding energies and spatial structures of
these complexes are presented as a function of the layer thickness.  The trion
and biexciton of the thinnest layered HOIP have binding energies of 35~meV and
44~meV, respectively, whereas a single exfoliated layer is predicted to have
trions and biexcitons with equal binding enegies of 48~meV.  We compare our
findings to available experimental data and to that of other
quasi-two-dimensional materials.
\end{abstract}

\maketitle

Layered hybrid organic-inorganic lead halide perovskites (HOIPs) have been
demonstrated to be a promising alternative material~\cite{tsai2016high, pedesseau2016advances}
to the three-dimensional (3D) HOIPs, which show remarkably high quantum yield and
power conversion efficiency~\cite{green2014emergence, saparov2016organic} but
are chemically unstable to air, moisture, and light~\cite{im20116, noh2013chemical, berhe2016organometal}.  
Layered HOIPs
such as (C$_4$H$_9$NH$_3$)$_2$(CH$_3$NH$_3$)$_{n-1}$Pb$_n$I$_{3n+1}$ consist of
quasi-two-dimensional quantum wells of lead halide octahedra that mimic the functional
capabilities of the 3D HOIPs, and spacers of bulky organic molecules that
provide the desired stability~\cite{zhang2017stable}.  Layered HOIPs are highly
tunable not only via the chemical composition, but also the thickness of the lead
halide quantum well, $n$, which can be controlled as a result of recent
synthetic advances~\cite{cao20152d, wu2015excitonic,
stoumpos2016ruddlesden, stoumpos2017high, weidman2017colloidal,
blancon2018scaling}.

Beyond their improved stability, the layered HOIPs also have potentially
different optoelectronic properties than the 3D HOIPs, and these differences
must be understood to enable technological applications.
In particular, quantum confinement and dielectric contrast arising from the organic spacer
increase the strength of the Coulomb interaction leading to the formation of tightly bound excitons
\cite{hong1992dielectric, tanaka2005image, shimizu2005influence,
katan2019quantum} and biexcitons~\cite{ishihara1992dielectric,
kato2003extremely, chong2016dominant, elkins2017biexciton, thouin2018stable},
which should be contrasted with the free charge carriers present in 3D HOIPs.
The similarity with other 2D semiconductors, such as transition-metal dichalcogenides (TMDCs),
suggests the possibility of trions~\cite{zahra2019screened} in the
presence of electron or hole doping; however, trions have not been experimentally reported
in layered HOIPs, to the best of our knowledge, perhaps due to difficulties in their doping.
Interestingly, the ordering of the binding energies of trions and biexcitons depends on
the interplay of dimensionality and dielectric contrast: in TMDCs the trion binding energy
is larger~\cite{Mayers2015,Zhang2015,Ye2018}, but in three-dimensional semiconductors 
and two-dimensional quantum wells without
dielectric contrast, the biexciton binding energy is larger~\cite{Usukura1999}.  Their ordering in
layered HOIPs has not been investigated until now.

These many-body excitonic physics of layered HOIPs are especially interesting
because of the strong electronic hybridization within layers, which makes their optoelectronic
properties more tunable than those of van der Waals materials.  However, this strong
hybridization also complicates the theoretical description of excitons, trions, and biexcitons
in layered HOIPs, as compared to the large body of work on strictly 2D materials.
In this paper, we develop a mixed effective-mass/tight-binding model
of excitons, trions, and biexcitons in layered HOIPs and we calculate their many-body
wave functions using newly developed extensions of an accurate variational method and 
diffusion Monte Carlo, which
have been the premier methods for analogous calculations on 
TMDCs~\cite{Ganchev2015,Mayers2015,Zhang2015,Kidd2016,Szyniszewski2017,Mostaani2017}.

\begin{figure}
    \centering
    \includegraphics{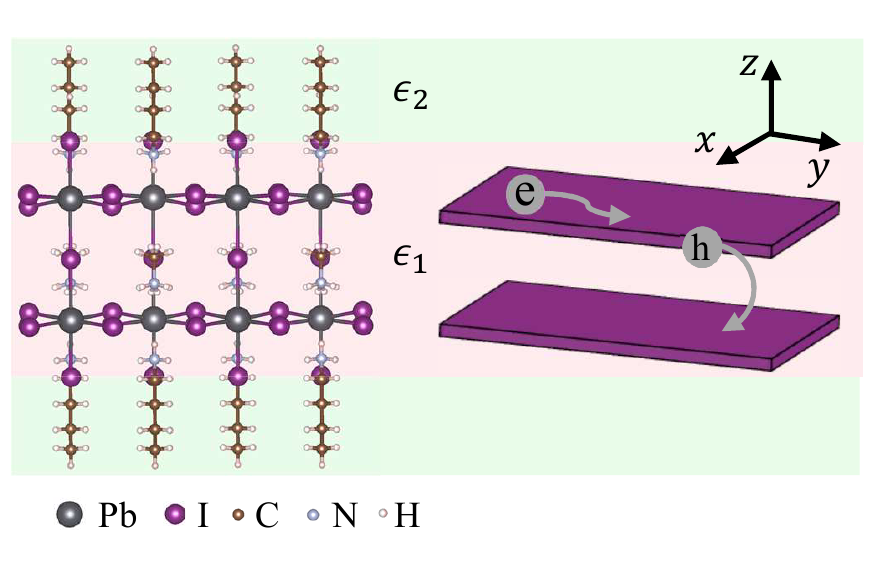}
    \caption{Illustration of our model of a layered HOIP, with $n=2$~\cite{stoumpos2016ruddlesden} shown 
as an example.  The electrons and holes move continuously in the $xy$ plane and hop discretely between
inorganic sublayers in the $z$ direction.  The electrostatics of the inorganic and organic layers are
modeled with slabs of dielectric constant $\epsilon_1$ and $\epsilon_2$, respectively.}
    \label{fig:model}
\end{figure}

As a prototypical system, we study
(C$_4$H$_9$NH$_3$)$_2$(CH$_3$NH$_3$)$_{n-1}$Pb$_n$I$_{3n+1}$.
One layer of Pb$_n$I$_{3n+1}$ is considered as $n$ discrete sublayers. Each
sublayer, which is a single layer of PbI$_4^{2-}$ octahedra, is treated as a
continuous two-dimensional plane and the particles are considered to be discretely hopping
between the sublayers, as shown in Figure~\ref{fig:model}.
The location of each particle $i$ is specified by its continuous position
in the $xy$ plane and a discrete sublayer position in the $z$ direction, 
i.e.~$(\bm{r}_i,z_i)$ where $z_i\in\{0, d, 2d, \ldots\}$ and $d$ is the sublayer spacing.
The total Hamiltonian for $N$ particles (electrons or holes) is
\begin{subequations}
\label{eq:ham}
\begin{align}
H &= \sum_{i=1}^{N} \big[ T_i + V(z_i) \big] + \sum_{i<j}^{N} q_i q_j W(\vr_i,z_i; \vr_j, z_j), \\
T_i &= -\frac{1}{2m_i}\nabla_{\vr_i}^2 - \sum_{z_i=0}^{(n-1)d} t_i (|z_i\rangle\langle z_i+d|+\mathrm{h.c.}), \\
V(z) &= \frac{1}{2}\lim_{\vr_1\rightarrow \vr_2} \left[ W(\vr_1,z;\vr_2,z) 
    - \frac{1}{\epsilon_1 |\vr_1-\vr_2|} \right],
\end{align}
\end{subequations}
and $q_1q_2W(\vr_1,z_1, \vr_2,z_2)$ is the screened Coulomb interaction
between charges $q_1$ and $q_2$ located at $(\vr_1,z_1)$ and $(\vr_2,z_2)$. To calculate
the screened Coulomb interaction, we use a model of infinitely alternating dielectric 
slabs~\cite{guseinov1984coulomb, muljarov1995excitons} as shown
in Figure~\ref{fig:model}.  The inorganic layer has dielectric constant 
$\epsilon_1 = 6.1$~\cite{ishihara1990optical} and thickness $n\times6.39$~\AA~\cite{oku2015crystal} and the organic
layer has dielectric constant $\epsilon_2 = 2.1$~\cite{ishihara1990optical} and thickness 8.81~\AA~\cite{muljarov1995excitons}.
The interaction $W$ can be straightforwardly obtained by classical electrostatics~\cite{guseinov1984coulomb, muljarov1995excitons}
and the one-body
potential $V(z)$ is the self-energy due to this dielectric
constrast~\cite{brus1983simple,cho2018environmentally}.
Based on the band structure of the cubic 3D HOIP~\cite{cho2019optical}, we use an effective mass $m_e = m_h = 0.2m_0$ for
the in-plane kinetic energy and a transfer integral $t = 1/(2md^2) = 0.47$~eV for the out-of-plane kinetic 
energy
where $d = 6.39$~\AA\ is the thickness of one sublayer.

The ground-state wave function of the Hamiltonian~(\ref{eq:ham}) is given by 
\begin{equation}
\label{eq:wfn}
    \ket{\Psi}=\sum_{\bm{z}}C_{\bm{z}}\ket{\psi_{\bm{z}}}\ket{\bm{z}},
\end{equation}
where $|\bm{z}\rangle = \prod_{i=1}^{N} |z_i\rangle$ indicates a configuration of sublayer positions
of each of the $N$ particles and $\psi_{\bm{z}}(\vr_1, \vr_2, \dots) \equiv \psi_{\bm{z}}(\vr)$ 
is the (normalized) in-plane 
wave function associated with that sublayer configuration.  For a layered HOIP with $N$ particles
in $n$ sublayers, there are $n^N$ configurations in the wave function expansion~(\ref{eq:wfn}), although
many are equivalent by spatial symmetry, which we exploit in our subsequent numerical studies.

We calculate the properties of excitons, trions, and biexcitons using two accurate, numerical
techniques: the stochastic variational method (SVM) with explicitly correlated
Gaussians (ECGs) and diffusion Monte Carlo (DMC).
In the SVM method, for each layer configuration we take a linear combination 
of ECGs~\cite{bubin2006matrix, mitroy2013theory}
\begin{subequations}
    \label{Eq:ECG}
\begin{align}
    |\psi_{\bm{z}}^{\mathrm{ECG}}\rangle &= \sum_{\alpha=1}^{N_\mathrm{G}}
        B_{\alpha}^{(\bm{z})} |\phi_{\alpha}^{(\bm{z})}\rangle \\
    \phi_{\alpha}^{(\bm{z})}(\vr) &=\sum_{\hat{P}}\hat{P}
        \exp\qty(-\sum_{i<j}^N [\mathbf{A}_\alpha^{(\bm{z})}]_{ij}(\bm{r}_i-\bm{r}_j)^2),
\end{align}
\end{subequations}
where $B_{\alpha}$ is an expansion coefficient to be variationally determined and $\hat{P}$ is a permutation
operator. The Gaussian parameters $A_{ij}$ are found by randomly generating 
multiple trial sets and keeping those that yield the lowest energy. 
For a given set of Gaussian parameters $A_{ij}$, we simultaneously optimize the linear expansion
coefficients $B_\alpha^{(\bm{z})}$ in Eq.~(\ref{Eq:ECG}) and 
$C_{\bm{z}}$ in Eq.~(\ref{eq:wfn}) by finding the ground-state eigenvector
of the generalized eigenvalue problem $\mathbf{HX} = \mathbf{SXE}$,
where matrices are expressed in the discrete space of Gaussians $\alpha$ and
layer configurations $\{\bm{z}\}$,
\begin{subequations}
\begin{align}
H_{\alpha\bm{z},\alpha^\prime\bm{z}^\prime} &= \int \prod_{i=1}^{N} d^2r_i\ \phi_\alpha^{(\bm{z})}(\vr)
    \langle \bm{z} | H | \bm{z^\prime}\rangle \phi_{\alpha^\prime}^{(\bm{z}^\prime)}(\vr) \\ 
S_{\alpha\bm{z},\alpha^\prime\bm{z}^\prime} &= \delta_{\bm{z},\bm{z}^\prime}
 \int \prod_{i=1}^{N} d^2r_i\ \phi_\alpha^{(\bm{z})}(\vr) 
    \phi_{\alpha^\prime}^{(\bm{z})}(\vr)
\end{align}
\end{subequations}
The variational energy is improved by increasing the number of ECGs
per sublayer configuration
$N_{\mathrm{G}}$. Here, convergence is determined to be achieved when the energy
difference is less than $10^{-3}$~meV for more than 3 subsequent SVM cycles.  We
address the increasing linear dependence of ECGs using canonical
orthogonalization~\cite{szabo2012modern}.

Here we also introduce an adaptation of the standard DMC method~\cite{Foulkes2001} that enables its application in a Hilbert space with both discrete and continuous components.
A wave function is evolved in imaginary time iteratively with a discrete time step $\Delta \tau$.
This function is represented by an ensemble of walkers, each with a position $(\bm{r}^{(w)}, \bm{s}^{(w)})$ and weight $x_w$ that evolve stochastically, where $\vr = (\vr_1,\ldots\vr_N)$ and $\bm{s}$ is a symmetry-adapted linear combination of $N$-particle configurations $\bm{z}$ of sublayer positions.
Each iteration involves three steps, the first and third of which simulate the action of the hopping terms in the Hamiltonian $H^{(z)} =  -\sum_{i=1}^N \sum_{z_i=0}^{(n-1)d} t_i \left( \ket{z_i}\bra{z_i + d} + \mathrm{h.c.} \right)$, and the second of which simulates the action of the remaining terms.
In the first and third steps, each walker is assigned a new discrete index $\bm{s}^{(w)\prime}$ sampled randomly from among all symmetry-adapted configurations $\lbrace \bm{s} \rbrace$, each with a probability proportional to $|Q_{\bm{s} \bm{s}^{(w)}}|$. 
The matrix $\mathbf{Q}$ is $\exp \left(-\Delta \tau \mathbf{J} / 2 \right)$, where  $\mathbf{J}$ is the matrix representation of $H^{(z)}$ in the basis of symmetry-adapted configurations.
The weight for each walker is multiplied by $\text{sgn}(Q_{\bm{s}^{(w)\prime} \bm{s}^{(w)}}) \sum_{\bm{s}} |Q_{\bm{s} \bm{s}^{(w)}}|$.

The second step involves displacing each walker from its position after the first step, $(\bm{r}^{(w)}, \bm{s}^{(w)\prime})$, to $(\bm{r}^{(w)\prime}, \bm{s}^{(w)\prime})$, where $\bm{r}^{(w)\prime}$ is sampled from a Gaussian distribution with variance $\Delta \tau$ and mean $[\bm{r}^{(w)} + \Delta \tau \bm{v}_d(\bm{r}^{(w)}, \bm{s}^{(w)\prime})]$.
The drift velocity $\bm{v}_d(\bm{r}, \bm{s}) = \nabla_{\bm{r}} \ln \Phi_g(\bm{r}, \bm{s})$, where $\Phi_g(\bm{r}, \bm{s})$ is a guiding wave function. 
Each walker's weight is multiplied by $\exp \left\{- \Delta \tau [E_L(\bm{r}^{(w)}, \bm{s}^{(w)}) + E_L(\bm{r}^{(w)\prime}, \bm{s}^{(w)\prime}) - 2 E_T] / 2 \right\}$, 
where the local energy is $E_L(\bm{r}, \bm{s}) = {\Phi_g(\bm{r}, \bm{s})}^{-1}\left(H - H^{(z)} \right) \Phi_g (\bm{r}, \bm{s})$.
The walkers' discrete indices are again updated, as in the first step, and 
the walker weights $\lbrace x_i \rbrace$ are resampled, such that some walkers are removed and others are duplicated. 
Ensemble energies calculated after each iteration are averaged to estimate the ground-state energy,
and extrapolation is used to reduce the time step error.
Further details of the SVM and DMC calculations, including the construction of symmetry-adapted configurations $\bm{s}$, are provided in the Supplemental Material.

\begin{figure}[t]
    \centering
    \includegraphics{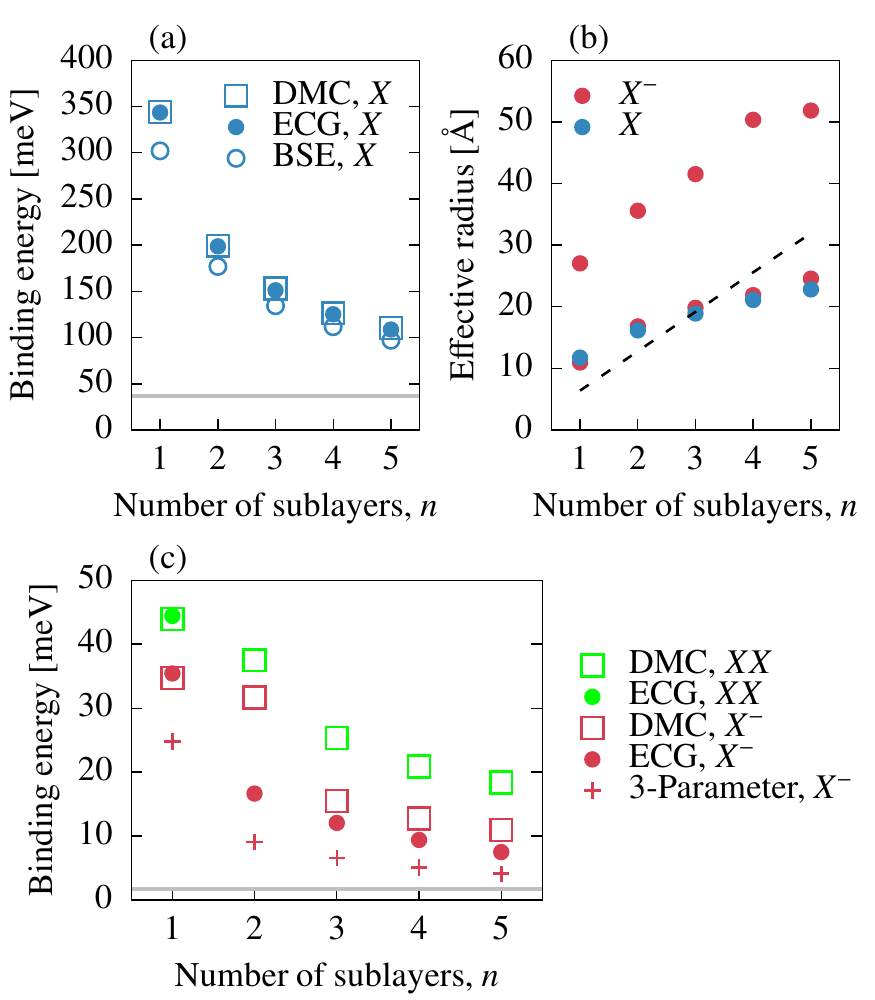}
    \caption{(a) Exciton binding energy of layered HOIPs with different number
of sublayers obtained by the SVM and DMC with the Hamiltonian~(\ref{eq:ham}) compared
with previously reported GW-BSE calculations with an atomistic tight-binding 
Hamiltonian~\cite{cho2019optical}.
(b) Effective radii of the exciton and trion as a function of number of
sublayers. Black dashed line indicates the total thickness of the quantum well,
$n\times d$. (c) Trion and biexciton binding energy calculated via DMC, SVM, and
the three-parameter variational wave function~(\ref{Eq:3param}).}
    \label{fig:tri_BE}
\end{figure}

\begin{figure}[t]
    \centering
    \includegraphics{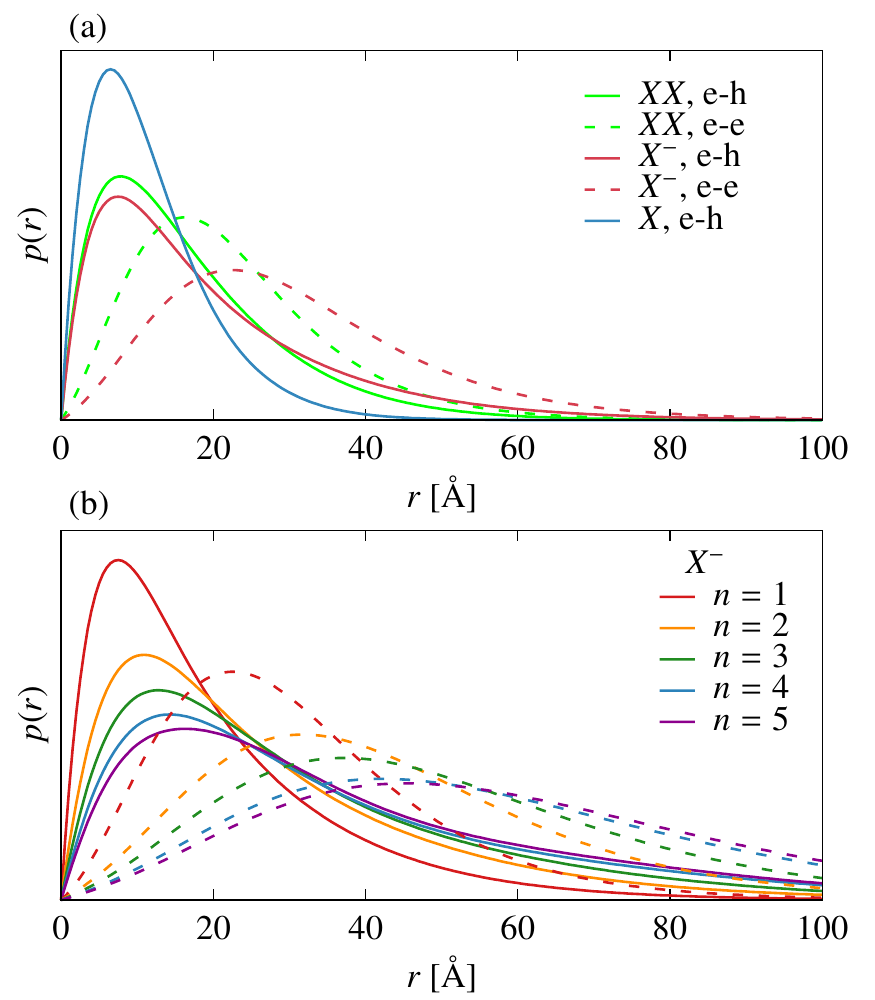}
    \caption{Radial probability distributions of electron-hole (solid line) and electron-electron (dashed line) distances of (a) exciton, trion, and biexciton of $n=1$ and (b) trion of $n=1-5$.}
    \label{fig:avg_r}
\end{figure}

The exciton properties (with $N=2$ particles) are straightforward to converge 
for both methods, which
agree with a numerically exact spatial grid-based diagonalization to within
2~meV.  Our exciton results allow us to assess the quality of the effective mass
+ tight-binding approximation in the Hamiltonian~(\ref{eq:ham}).  In
Fig.~\ref{fig:tri_BE}(a), we compare the obtained $n$-dependent exciton binding
energy (defined as the difference in energies between the ground states of the
Hamiltonian with and without the Coulomb interaction $W$) to that obtained by
two of us (Y.C.~and T.C.B.) in a previous atomistic calculation, which included
multiple bands with nonparabolicity~\cite{cho2019optical}.
We see that the current Hamiltonian overestimates the exciton binding energy by 10-40~meV,
an acceptable error of about 10\%.
These large exciton binding energies of 100-400~meV are comparable to those of
TMDCs~\cite{Chernikov2014,He2014,Ugeda2014}.

\begin{figure}[t!]
    \centering
    \includegraphics{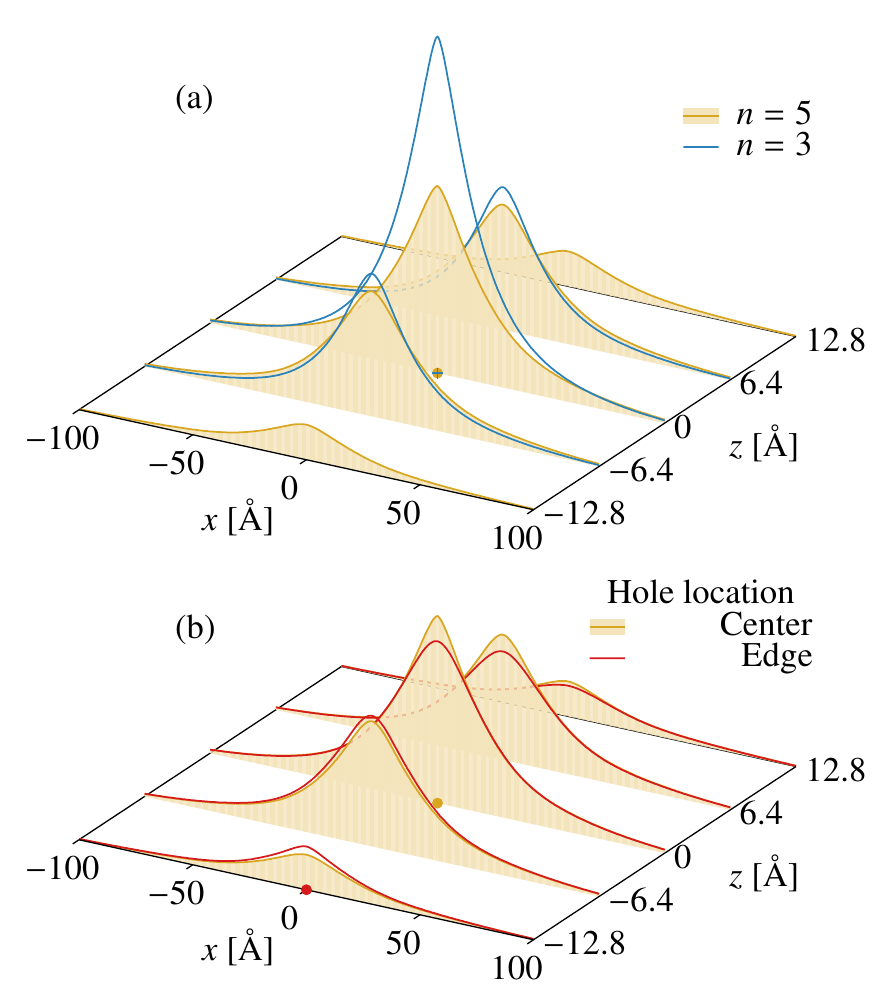}
    \caption{Total electron probability distribution of the trion for (a) $n=3$ and
$n=5$ with the hole fixed in the center sublayer, and (b) $n=5$ with the hole fixed at
the center and the edge sublayers.}
    \label{fig:n5}
\end{figure}

The binding energies of trions ($N=3$) and biexcitons ($N=4$) calculated using the SVM
and DMC are presented in Figure~\ref{fig:tri_BE}(c) as a function of the
number of sublayers $n$. Like the exciton,
the trion and biexciton have very large binding energies of 10-50~meV, which is again
similar to that in TMDCs~\cite{Mak2013,Berkelbach2013,Mayers2015,Zhang2015,Ganchev2015,Kidd2016,Mostaani2017}.  
Focusing first on the trion, we see good agreement between
the SVM and DMC results except for $n=2$, for which the binding energies differ by about
15~meV (a factor of 2); this discrepancy may be a combination of problematic
convergence in the SVM and time step error in the DMC.  For $n=1$ and $n\ge 3$, the trion
binding energies agree to within 4~meV.
For the trion (assumed negative), we also consider a simple form of in-plane
trial wave function $\psi_{\bm{z}}$ that has three variational parameters for
each sublayer configuration 
\begin{equation}
    \psi_{\bm{z}}^{(3)} \propto [\exp(-\rho_1/a^{(\bm{z})}-\rho_2/b^{(\bm{z})})+(1\leftrightarrow 2)]
    (1+c^{(\bm{z})}\rho_{12}),
    \label{Eq:3param}
\end{equation}
where $\rho_1$ and $\rho_2$ are the in-plane separations of each electron with
the hole and $\rho_{12}$ is the in-plane separation of the two
electrons~\cite{chandrasekhar1944some}. 

While the binding energy of the three-parameter wave function only recovers about 50-70\% of the 
SVM binding energy, as shown in Figure \ref{fig:tri_BE}(c), it provides a useful physical picture 
through the parameters $a$ and $b$, which can be interpreted as the effective radius of the two electrons.
Figure~\ref{fig:tri_BE}(b) compares the effective in-plane radii of the exciton and trion. 
The effective radius of the exciton is the average of the electron-hole distance. The 
effective radii of the trion are those that best fit the average in-plane electron-hole 
radial probability distribution of the optimized SVM wave function. 
The smaller radius $a$ is remarkably similar to that of the exciton and also comprable to the total
thickness of the inorganic quantum well.  The larger radius $b$ is about twice $a$, indicating that the
``second'' electron is weakly bound.  

For the $n=1$ biexciton, SVM and DMC are in excellent agreement and both methods
predict a biexciton binding energy of about 44~meV, which is in agreement with
the experimentally observed value of 
40-50~meV~\cite{ishihara1992dielectric, chong2016dominant, thouin2018stable}
and about twice as large as in monolayer TMDCs~\cite{Mayers2015,Zhang2015}. 
Moreover, the calculations show that the biexciton is more strongly bound than the
trion by almost 10~meV -- an energetic ordering that is
\textit{opposite} to that observed in monolayer TMDCs~\cite{Mayers2015,Zhang2015,Ye2018}, as discussed in the
introduction.  We have confirmed that this difference is due to the smaller
dielectric contrast of the perovskites.  The aforementioned convergence
problems of the SVM were especially severe for the $n>1$ biexciton because of the exponentially large
number of variational paramers.  However, our DMC calculations demonstrate that
this energetic ordering is maintained for larger values of $n$, consistent with
the known ordering of trions and biexcitons in 3D systems~\cite{Usukura1999}.
The bulk 3D values can be simply estimated using the relevant Rydberg energy scale
and previous numerical results~\cite{Usukura1999}, leading to a biexciton binding energy of
2.3~meV \cite{kylanpaa2009thermal} and a trion binding energy of 1.8~meV
\cite{kar2006positron}.  Therefore, even for $n=5$, for which the biexciton and
trion binding energies are about 18~meV and 11~meV, the excitonic complexes are
still sensitive to electronic and dielectric contrast provided by the organic
layers.

The spatial structures of the exciton, trion, and biexciton provide further insight.
In Figure~\ref{fig:avg_r}(a), we show the electron-hole and electron-electron 
in-plane radial probability distributions for the $n=1$ exciton, trion, and biexciton,
as calculated by the SVM.
We see that the electron-hole distribution is similar for all three complexes,
whereas the electron-electron distribution is quite different between the trion
and biexciton.  In Figure~\ref{fig:avg_r}(b), we extend this analysis to
$n\ge1$, focusing on the spatial structure of the trion.  For $n>1$, we
calculate the radial probability distributions as the expectation over all layer
configurations.  Consistent with our simple picture based on the three-parameter
wave function, we see that the trion becomes signifantly larger for increasing $n$,
with tails extending beyond~100~\AA. 

\setlength{\tabcolsep}{10pt}
\renewcommand{\arraystretch}{1.5}
\begin{table}[t!]
\caption{Binding energy (in meV) of the exciton, trion, and biexciton of an infinite
(bulk) layered HOIP and exfoliated single-layer HOIP, both with $n=1$. The exfoliated sample is assumed to be
suspended in vacuum and have one formula unit of (C$_4$H$_9$NH$_3$)$_2$PbI$_4$.}
\label{tab:exfol} 
\begin{tabular}{p{.14\textwidth}ccc}
\hline\hline
$n=1$ & Exciton & Trion & Biexciton \\ \hline 
Infinite (bulk)      & 344     & 35.4  & 44.3      \\ \hline
Exfoliated           & 662     & 48.5  & 48.5      \\ \hline \hline
\end{tabular}
\end{table}

We next aim to quantify the electron-hole correlation in the vertical direction
and its sensitivity to the vertical confinement, focusing again on the trion.
In Figure~\ref{fig:n5}(a) we plot the total probability distribution of the electrons
when the hole is located in the center layer of the $n=3$ and $n=5$ HOIPs.
The weaker confinement of $n=5$ 
reduces the short-range electron-hole correlation, while leaving the long-range
correlation relatively constant. 
In Figure~\ref{fig:n5}(b), we compare the electron distributions of the $n=5$ trion
when the hole is in the center and when the hole is at the edge.  Again,
only the short-range correlation is significantly impacted.  Moreoever, we see that
even when the hole is fixed to the edge of the inorganic layer, the electron density is
still maximal in the center of the layer.  This result indicates that the combination of
carrier confinement and repulsive electron-electron interactions dominate the
attractive electron-hole interaction.

Finally, we consider the excitonic properties of a hypothetical exfoliated $n=1$
perovskite layer (or nanoplatelet), motivated by a number of experimental
efforts~\cite{niu2014exfoliation, yaffe2015excitons, dou2015atomically}.
This system is especially interesting for a number of reasons.  First, the exfoliation
enhances the dielectric contrast, increasing the strength of the Coulomb interaction
and making the behavior more similar to that of monolayer TMDCs.  Second, such an atomically
thin sample could potentially be electrostatically gated in order to induce carrier doping.
Therefore, the experimental observation of trions in HOIPs may be easiest for exfoliated
samples.  
To model the screened Coulomb interaction $W$ of an exfoliated HOIP layer, we use the
electrostatic transfer matrix method~\cite{cavalcante2018electrostatics} with five regions:
vacuum, organic, inorganic, organic, and vacuum.
In Table~\ref{tab:exfol}, we present the exciton, trion, and biexciton binding
energies of the infinite and exfoliated $n=1$ HOIPs, as predicted by the SVM.
Unsurprisingly, the exciton has an extremely large binding energy of 662~meV.  
Remarkably, the trion and biexciton have \textit{identical} binding energies.
Therefore, this system is right on the cusp of the crossover between
a 2D material with strong dielectric contrast (for which the trion binding energy
is largest) and a 3D material without dielectric contrast (for which the
biexciton binding energy is largest).  Although the similarity in binding energies may
obfuscate their experimental identification, it should be possible to distinguish them
through a study of the doping and excitation fluence dependence.

To summarize, we have proposed and parameterized a microscopic model Hamiltonian 
for excitonic complexes in layered HOIPs.
With this Hamiltonian, we have studied the properties of
excitons, trions, and biexcitons using carefully converged 
SVM and DMC calculations, finding large binding energies that are similar
to those observed in TMDCs. Our methods can be readily applied to other
layered and mixed-dimensional materials, and can be extended to
describe the recently reported photonic behavior of anisotropic excitons in
layered HOIPs~\cite{Guo2018}.
We hope that our
calculations, demonstrating the
existence of stable trions with large binding energies, motivates efforts to dope
HOIPs in search of trions.  In particular, we imagine that the properties of
doped HOIPs will shed light on their high carrier mobilities and defect tolerance~\cite{Kang2017}.
Finally, we propose that doped, layered HOIPs may be a suitable platform to study
the crossover from two-dimensional to three-dimensional
exciton-polarons~\cite{Efimkin2017a,Sidler2017}; however, the difficulty in
doping may preclude the sufficiently high carrier densities needed to observe
exciton-polaron physics.

\vspace{1em}

T.C.B.~thanks Alexey Chernikov and Omer Yaffe for useful discussions. 
This work was supported in part by the Air Force Office of Scientific Research
under AFOSR Award No.~FA9550-19-1-0405 (Y.C.~and T.C.B.) and by a software fellowship from the
Molecular Sciences Software Institute, which is funded by U.S. National Science
Foundation grant OAC-1547580 (S.M.G.).  We acknowledge computing resources from
Columbia University’s Shared Research Computing Facility project, which is
supported by NIH Research Facility Improvement Grant 1G20RR030893-01, and
associated funds from the New York State Empire State Development, Division of
Science Technology and Innovation (NYSTAR) Contract C090171, both awarded April
15, 2010.  The Flatiron Institute is a division of the Simons Foundation.



\end{document}


\title{Supplemental Material: Simulations of Trions and Biexcitons in Layered Hybrid Organic-Inorganic Lead Halide Perovskites}

\author{Yeongsu Cho}
\author{Samuel M. Greene}
\affiliation{Department of Chemistry, 
Columbia University, New York, New York 10027, USA}
\author{Timothy C. Berkelbach}
\affiliation{Department of Chemistry, 
Columbia University, New York, New York 10027, USA}
\affiliation{Center for Computational Quantum Physics, Flatiron Institute, New York, New York 10010, USA}

\maketitle

\renewcommand\theequation{S\arabic{equation}}
\renewcommand\thesection{S\arabic{section}}
\renewcommand\thefigure{S\arabic{figure}}

\section{Constructing a Symmetry-Adapted Discrete Basis}
\label{sec:symm}
Many of the discrete configurations $\lbrace \bm{z} \rbrace$ for a layered HOIP are symmetrically equivalent.
Two configurations are equivalent if they are related by a reflection of all particles across a plane in the center of a sublayer, by an exchange of all holes for electrons and all electrons for holes, or by composition of both operations.
This symmetry can be leveraged to reduce the dimension of the discrete part of Hilbert space, thereby enabling improved computational efficiency in calculations.
For ground-state calculations, the unitary transformation matrix $\mathbf{U}$ relating the $n^N$ discrete configurations $\lbrace \bm{z} \rbrace$ to a smaller set of symmetry-adapted configurations $\lbrace \bm{s} \rbrace$, i.e.
\begin{equation}
\ket{\bm{s}} = \sum_{\bm{z}} U_{\bm{z} \bm{s}} \ket{\bm{z}}
\end{equation}
can be calculated straightforwardly as follows.
If $\mathcal{C}^{(\bm{s})}$ denotes the subset of all configurations $\lbrace \bm{z} \rbrace$ associated with a single symmetry-adapted configuration $\bm{s}$, then elements in the corresponding column of $\mathbf{U}$ are defined as
\begin{equation}
U_{\bm{z} \bm{s}} = 
\begin{cases}
|\mathcal{C}^{(\bm{s})}|^{-1/2} & \bm{z} \in \mathcal{C}^{(\bm{s})} \\
0 & \bm{z} \notin \mathcal{C}^{(\bm{s})} \\
\end{cases}
\end{equation}
Note that all such subsets are disjoint.
Any matrix or operator $\mathbf{A}$ defined in the basis of all configurations $\lbrace \bm{z} \rbrace$ can be transformed into this basis of symmetry-adapted configurations as $\mathbf{U}^\text{T}\mathbf{AU}$.

An analogous technique is used to reduce the dimensions of the $\mathbf{H}$ and $\mathbf{S}$ matrices used to obtain the ground-state eigenvector $\mathbf{x}_0$ of the generalized eigenvalue problem $\mathbf{Hx}^{(0)} = E^{(0)} \mathbf{Sx}^{(0)}$ in the SVM.
The $\mathbf{H}$ and $\mathbf{S}$ matrices are transformed as
\begin{equation}
\tilde{H}_{\alpha\bm{s},\alpha^\prime\bm{s}^\prime} = \sum_{\bm{z}, \bm{z}'} U_{\bm{z}, \bm{s}} H_{\alpha\bm{z},\alpha^\prime\bm{z}^\prime} U_{\bm{z}', \bm{s}'}
\end{equation}
and
\begin{equation}
\tilde{S}_{\alpha\bm{s},\alpha^\prime\bm{s}^\prime} = \sum_{\bm{z}, \bm{z}'} U_{\bm{z}, \bm{s}} S_{\alpha\bm{z},\alpha^\prime\bm{z}^\prime} U_{\bm{z}', \bm{s}'}
\end{equation}
The ground-state eigenvector $\tilde{\mathbf{x}}^{(0)}$ of the transformed eigenvalue problem $\tilde{\mathbf{H}}\tilde{\mathbf{x}}^{(0)} = E^{(0)} \tilde{\mathbf{S}}\tilde{\mathbf{x}}^{(0)}$ is related to that of the original eigenvalue problem by
\begin{equation}
x^{(0)}_{\alpha\bm{z}} = \sum_{\bm{s}} U_{\bm{z}\bm{s}} \tilde{x}^{(0)}_{\alpha \bm{s}}
\end{equation}

\section{Diffusion Monte Carlo in mixed discrete/continuous Hilbert space}
The Diffusion Monte Carlo (DMC) algorithm~\cite{Foulkes2001} involves propagating an initial state in imaginary time and applying Monte Carlo sampling to control the computational complexity of this propagation. This section describes how we modified the standard real-space formulation of DMC to enable its application in a Hilbert space with both discrete and continuous components. First, we describe the modifications to the differential equations and Green's functions used to propagate a generic state, as dictated by the form of the Hamiltonian in the main text. We then describe a specific wave function ansatz suitable for application to a mixed discrete/continuous Hilbert space (based on time-evolving walkers, as in standard DMC). The sampling techniques used to propagate this ansatz are discussed. Finally, we discuss some implementation details for the calculations presented in this work.

\subsection{Differential Equations for Imaginary Time Propagation}
An arbitrary quantum state for a system in a space with both discrete and continuous components can be expressed as
\begin{equation}
\ket{\Psi(\tau)} = \sum_{\bm{s}} \int \text{d}\bm{r} \psi(\bm{r}, \bm{s}, \tau) \ket{\bm{r}} \ket{\bm{s}},
\end{equation}
where $\tau$ is an index that here denotes imaginary time, $\bm{r}$ is a vector specifying the positions of each of the particles in continuous space $\left( r_1, r_2, ..., r_N \right)$, and $\bm{s}$ is a symmetry-adapted linear combination of configurations.
This state is propagated according to the imaginary-time Schr\"odinger equation:
\begin{equation}
\label{eq:tdSchro}
-\frac{\partial}{\partial \tau} \ket{\Psi(\tau)} = \left(H - E_T(\tau) \right) \ket{\Psi(\tau)}
\end{equation}
where $E_T(\tau)$ is an energy shift adjusted dynamically to ensure that the norm of the solution remains finite as it is propagated to longer times.
The method described here can be applied when the Hamiltonian is of the form
\begin{equation}
\label{eq:ham}
H = H^{(r)} + H^{(z)}
\end{equation}
where $H^{(r)}$ connects Hilbert space basis elements that differ only in their continuous component, and $H^{(z)}$ connects those that differ only in their discrete component. For the Hamiltonian in the main text,
\begin{equation}
H^{(r)} = \sum_{i=1}^N \left( -\frac{1}{2 m_i} \nabla^2_{\bm{r}_i} + V(z_i) \right) + \sum_{i < j} W(\bm{r}_i, z_i; \bm{r}_j, z_j)
\end{equation}
and
\begin{equation}
H^{(z)} = \sum_{i=1}^N \sum_{z_i=1}^{n-1} -t_i \left( \ket{z_i}\bra{z_i + 1} + \ket{z_i + 1}\bra{z_i} \right)
\end{equation}
In this case, the system of coupled differential equations equivalent to (\ref{eq:tdSchro}), one for each symmetry-adapted configuration $\bm{s}$ in discrete space, is
\begin{equation}
-\frac{\partial}{\partial \tau} \psi(\bm{r}, \bm{s}, \tau) = \left( H^{(r)} - E_T(\tau) \right) \psi(\bm{r}, \bm{s}, \tau) + \sum_{\bm{s}'} T_{\bm{s}, \bm{s}'} \psi(\bm{r}, \bm{s}', \tau)
\end{equation}
where the sum is over all discrete states $\bm{s}'$.
Here, $\mathbf{T}$ is the matrix representation of $H^{(z)}$ in the symmetry-adapted discrete basis, obtained by applying the transformation $\mathbf{U}$ to the matrix representation of $H^{(z)}$ in the basis of all configurations, i.e.with elements $\mel{\bm{z}}{H^{(z)}}{\bm{z}'}$.
In order to reduce the statistical error arising from the Monte Carlo sampling used to solve this system of differential equations, each equation is multiplied by a guiding function $\Phi_g(\bm{r}, \bm{s})$, in analogy to standard importance-sampled DMC:
\begin{equation}
-\frac{\partial}{\partial \tau} f(\bm{r}, \bm{s}, \tau) = \Phi_g(\bm{r}, \bm{s}) \left( H^{(r)} - E_T(\tau) \right) \left[\frac{f(\bm{r}, \bm{s}, \tau)}{\Phi_g(\bm{r}, \bm{s})} \right] + \Phi_g(\bm{r}, \bm{s}) \sum_{\bm{s}'} T_{\bm{s}, \bm{s}'} \frac{f(\bm{r}, \bm{s}', \tau)}{\Phi_g(\bm{r}, \bm{s}')}
\end{equation}
The function $f(\bm{r}, \bm{s}, \tau)$ is defined as $\psi(\bm{r}, \bm{s}, \tau) \Phi_g(\bm{r}, \bm{s})$.

The solution to this differential equation, $f(\bm{r}, \bm{s}, \tau)$, can be approximated by standard operator splitting methods~\cite{McLachlan2002, Splitting2016} (a generalization of the Trotter-Suzuki expansion~\cite{Trotter1959, Suzuki1977}), solving each of the following two systems of differential equations in turn, each with a finite time step. 
The first involves the Hamiltonian component that connects elements in discrete space:
\begin{equation}
\label{eq:hopDifEq}
-\frac{\partial}{\partial \tau} f(\bm{r}, \bm{s}, \tau) = \Phi_g(\bm{r}, \bm{s}) \sum_{\bm{s}'} T_{\bm{s}, \bm{s}'} \frac{f(\bm{r}, \bm{s}', \tau)}{\Phi_g(\bm{r}, \bm{s}')}
\end{equation}
and the second involves the continuous-space component:
\begin{equation}
\label{eq:xyDifEq}
-\frac{\partial}{\partial \tau} f(\bm{r}, \bm{s}, \tau) = \Phi_g(\bm{r}, \bm{s}) \left( H^{(r)} - E_T(\tau) \right) \left[\frac{f(\bm{r}, \bm{s}, \tau)}{\Phi_g(\bm{r}, \bm{s})} \right]
\end{equation}
If this is done by propagating (\ref{eq:hopDifEq}) with a time step of $\Delta \tau / 2$, followed by (\ref{eq:xyDifEq}) with a time step of $\Delta \tau$, then (\ref{eq:hopDifEq}) with a time step of $\Delta \tau / 2$, the resulting discretization error is $\mathcal{O}(\Delta \tau^3)$~\cite{Splitting2016}.
If (\ref{eq:hopDifEq}) is approximated as
\begin{equation}
\label{eq:approxHop}
-\frac{\partial}{\partial \tau} f(\bm{r}, \bm{s}, \tau) \approx \sum_{\bm{s}'} T_{{\bm{s}, \bm{s}'}} f(\bm{r}, \bm{s}', \tau)
\end{equation}
its solution can be simplified. It is given as
\begin{equation}
\label{eq:hopGreenfxn}
f(\bm{r}, \bm{s}, \tau + \Delta \tau/ 2) = \sum_{\bm{s}'} M_{\bm{s}, \bm{s}'} f(\bm{r}, \bm{s}', \tau)
\end{equation}
where the matrix $\mathbf{M} = \mathbf{U}^\text{T} \exp(- \Delta \tau \mathbf{T} / 2) \mathbf{U}$ is calculated using standard numerical matrix exponentiation tools. This approximation (\ref{eq:approxHop}) becomes exact as $\Delta \tau \to 0$ or $\mathbf{T} \to 0$, or as $\Phi_g(\bm{r}, \bm{s})$ becomes independent of $\bm{s}$.
The severity of this approximation for a particular Hamiltonian and choice of $\Phi_g(\bm{r}, \bm{s})$ can be monitored by observing the convergence of the solution as $\Delta \tau \to 0$.
The solution to (\ref{eq:xyDifEq}) is given in analogy to standard DMC as
\begin{equation}
\label{eq:ctsGreenfxn}
f(\bm{r}, \bm{s}', \tau + \Delta \tau) = \int \text{d}\bm{r} \tilde{G}(\bm{r} \to \bm{r}', \bm{s}, \Delta \tau) f(\bm{r}, \bm{s}, \tau)
\end{equation}
where $\tilde{G}(\bm{r} \to \bm{r}', \bm{s}, \Delta \tau)$ is approximated as a product of a drift-diffusion term and a branching term, $\tilde{G}_d(\bm{r} \to \bm{r}', \bm{s}, \Delta \tau) \tilde{G}_b(\bm{r} \to \bm{r}', \bm{s}, \Delta \tau)$, where
\begin{equation}
\label{eq:driftdifGF}
\tilde{G}_d(\bm{r} \to \bm{r}', \bm{s}, \Delta \tau) = (2 \pi \Delta \tau)^{-D/2} \exp \left[ -\frac{\left( \bm{r}' - \bm{r} - \Delta \tau \bm{v}_d(\bm{r}, \bm{s}) \right)^2}{2 \Delta \tau} \right]
\end{equation}
Here $D$ is the dimension of the continuous component of Hilbert space ($2N$ for the systems considered here), and $\bm{v}_d(\bm{r}, \bm{s})$ is a drift velocity, defined as
\begin{equation}
\bm{v}_d(\bm{r}, \bm{s}) = \nabla_{\bm{r}} \ln \left\lvert \Phi_g(\bm{r}, \bm{s}) \right\rvert
\end{equation}
The branching term is defined as
\begin{equation}
\label{eq:branchGF}
\tilde{G}_b(\bm{r} \to \bm{r}', \bm{s}, \Delta \tau) = \exp \left[ - \Delta \tau \left(E_L(\bm{r}, \bm{s}) + E_L(\bm{r}, \bm{s}') - 2 E_T(\tau) \right) / 2 \right]
\end{equation}
where $E_L(\bm{r}, \bm{s})$ is the local energy:
\begin{equation}
E_L(\bm{r}, \bm{s}) = \frac{H^{(r)} \Phi_g(\bm{r}, \bm{s})}{\Phi_g(\bm{r}, \bm{s})}
\end{equation}

\subsection{Monte Carlo Sampling}
Having introduced the differential equations describing imaginary-time propagation, we next describe their application to the specific ansatz considered here.
The function $f(\bm{r}, \bm{s}, \tau)$ at each iteration $\tau$ is represented by an ensemble of $N_w$ walkers, each at a specific position $( \bm{r}^{(i)}, \bm{s}^{(i)})$ in Hilbert space:
\begin{equation}
\label{eq:DMCwf}
f(\bm{r}, \bm{s}, \tau) = \sum_i^{N_w} x_i \delta_{\bm{s}, \bm{s}^{(i)}} \delta(\bm{r} - \bm{r}^{(i)})
\end{equation}
Here $x_i$ denotes the weight associated with each walker, which is allowed to be positive or negative.
This function is propagated according to (\ref{eq:approxHop}) and (\ref{eq:xyDifEq}), and Monte Carlo sampling is applied after each propagation step to ensure that the solution maintains this functional form.
Applying the propagation step involving the discrete Hamiltonian component (\ref{eq:approxHop}) involves updating the discrete index of each walker in discrete space $\bm{s}^{(i)}$ to a new index $\bm{s}^{(i)\prime}$, where $\bm{s}^{(i)\prime}$ is sampled randomly from the discrete probability distribution
\begin{equation}
\label{eq:discrProb}
p(\bm{s} | \bm{s}^{(i)}) = \frac{\left\lvert M_{\bm{s}, \bm{s}^{(i)}} \right\rvert}{\sum_{\bm{s}'} \left\lvert M_{\bm{s}', \bm{s}^{(i)}} \right\rvert}
\end{equation}
After the random selection of a specific index $\bm{s}^{(i)\prime}$ for each walker, its weight is updated. Its updated weight depends on the selected index as follows:
\begin{equation}
\label{eq:hopWeight}
x_i'(\bm{s}^{(i)\prime}) = x_i \text{sgn}[M_{\bm{s}^{(i)\prime}, \bm{s}^{(i)}}] \sum_{\bm{s}'} \left\lvert M_{\bm{s}', \bm{s}^{(i)}} \right\rvert
\end{equation}
The resulting solution after this propagation step is
\begin{equation}
f(\bm{r}, \bm{s}, \tau + \Delta \tau/ 2) = \sum_i^{N_w} x_i'(\bm{s}^{(i)\prime}) \delta_{\bm{s}, \bm{s}^{(i)\prime}} \delta(\bm{r} - \bm{r}^{(i)})
\end{equation}
In order to understand why this result is equal in expectation to the function obtained without random sampling, i.e. by substituting (\ref{eq:DMCwf}) into (\ref{eq:hopGreenfxn})
\begin{equation}
\label{eq:hopNoSamp}
f(\bm{r}, \bm{s}, \tau + \Delta \tau/ 2) = \sum_i^{N_w} x_i \delta(\bm{r} - \bm{r}^{(i)}) \sum_{\bm{s}'} M_{\bm{s}^{(i)}, \bm{s}'} \delta_{\bm{s},\bm{s}'}
\end{equation}
one can consider summing over all possible sampling outcomes for each walker and multiplying the result of each outcome by its corresponding probability
\begin{equation}
f(\bm{r}, \bm{s}, \tau + \Delta \tau/ 2) = \sum_i^{N_w} \delta(\bm{r} - \bm{r}^{(i)}) \sum_{\bm{s}'} p(\bm{s}' | \bm{s}^{(i)}) x_i'(\bm{s}') \delta_{\bm{s}, \bm{s}'}
\end{equation}
Substituting (\ref{eq:discrProb}) and (\ref{eq:hopWeight}) into this expression yields (\ref{eq:hopNoSamp}).

In each iteration, after the solution is propagated according to (\ref{eq:hopGreenfxn}), it is then propagated according to (\ref{eq:ctsGreenfxn}) as in standard real-space DMC. 
First, a new position for each walker is sampled randomly according to the Gaussian distribution in (\ref{eq:driftdifGF}). 
Some proposed moves are rejected according to the algorithm described in ref \citenum{Umrigar1993} in order to reduce the error resulting from a finite time step $\Delta \tau$. 
The weight of each walker is then multiplied by the branching term (\ref{eq:branchGF}).

After the second propagation step involving the discrete part of the Hamiltonian, the numerator and denominator of an energy estimator, $E(\tau) = E_n(\tau) / E_d(\tau)$, are calculated as
\begin{equation}
\label{eq:enNum}
E_n(\tau) = \mel{\Phi_T}{H}{\Psi(\tau)}
\end{equation}
and
\begin{equation}
\label{eq:enDen}
E_d(\tau) = \braket{\Phi_T}{\Psi(\tau)}
\end{equation}
where $\ket{\Phi_T}$ is a trial state chosen to approximate the ground state, and $\ket{\Psi(\tau)}$ is related to the current ensemble of walkers by
\begin{equation}
\ket{\Psi(\tau)} = \sum_{\bm{s}} \int \text{d}\bm{r} \frac{f(\bm{r}, \bm{s}, \tau)}{\Phi_g(\bm{r}, \bm{s})} \ket{\bm{r}} \ket{\bm{s}}
\end{equation}
The form of $\ket{\Phi_T}$ used in this work will be discussed below. 
The ground-state energy is estimated as the quotient of the average values of $E_n(\tau)$ and $E_d(\tau)$. 
Typically, iterations before a suitably chosen burn-in time are excluded from these averages~\cite{Chodera2016}. 
The numerator and denominator are averaged separately to avoid introducing an additional statistical bias~\cite{Blunt2014, Blunt2019}.

The final step in each iteration involves redistributing the weights of all walkers randomly according to a resampling procedure. 
Here we use a systematic sampling procedure~\cite{Lim2017} instead of the binomial procedure applied more commonly in DMC~\cite{Foulkes2001}, as the systematic procedure is known to yield less sampling error in some cases. 
This is not dictated by the adaptation of the DMC algorithm for a mixed discrete/continuous Hilbert space, and the binomial procedure could be straightforwardly applied instead.
The systematic procedure involves sampling a single random number $r$ uniformly from the interval $[0, 1)$. 
Then, for each walker $i$, a nonnegative integer $n^{(i)}$ is calculated, where $n^{(i)}$ is the number of integer values of $k$ that satisfy
\begin{equation}
\sum_{j=1}^{i - 1} |x_j| \leq \frac{k + r}{N_w}\sum_{l=1}^{N_w} |x_l| < \sum_{m=1}^{i} |x_m|
\end{equation}
Each walker is replaced by $n^{(i)}$ copies of itself; walkers for which $n^{(i)} = 0$ are removed from the simulation.
The weight of each walker is then updated to
\begin{equation}
\frac{\text{sgn}(x_i)}{N_w} \sum_{l=1}^{N_w} |x_l|
\end{equation}
such that the magnitudes of all weights are now equal. A benefit of this approach is that the number of walkers $N_w$ remains constant as the simulation proceeds. The energy shift $E_T(\tau)$ is then adjusted as
\begin{equation}
\label{eq:shiftUpdate}
E_T(\tau + \Delta \tau) = E_T(\tau) - \frac{\xi}{\Delta \tau} \ln \frac{S(\tau)}{S(\tau - \Delta \tau)}
\end{equation}
where $\xi$ is a damping parameter (chosen to be 0.05 in this work), and $S(\tau)$ denotes the sum of the magnitudes of walker weights at iteration $\tau$.

\subsection{Implementation Details}
We next discuss the specific choices made in implementing the method described above in this work. A product exponential function was used as the guiding wave function:
\begin{equation}
\Phi_g(\bm{r}, \bm{s}) = \Phi_g(\bm{r}, \bm{z}) = \prod_{i<j}^N \exp \left[a_{ij} \left(||\bm{r}_i - \bm{r}_j||_2^2 + (z_i - z_j)^2 \right)^{1/2} \right]
\end{equation}
where $\bm{z}$ denotes any one of the symmetrically equivalent configurations in $\mathcal{C}^{(s)}$ (Section~\ref{sec:symm}).
The $a_{ij}$ parameters are chosen to satisfy the appropriate cusp conditions:
\begin{equation}
a_{ij} = \begin{cases}
-2\mu_{ij} / \epsilon_1 & q_i q_j = 1 \\
+2\mu_{ij} / \epsilon_1 & q_i q_j = -1
\end{cases}
\end{equation}
where $q_i$ represents the charge of particle $i$, $\mu_{ij}$ represents a reduced mass, and $\epsilon_1$ is the dielectric constant defined in the main text.
This function was also used to define the trial state
\begin{equation}
\ket{\Psi_T} = \sum_{\bm{s}} c_{\bm{s}} \int \text{d} \bm{r} \Phi_g(\bm{r}, \bm{s}) \ket{\bm{r}} \ket{\bm{s}}
\end{equation}
as in this case the numerator and denominator of the energy estimator (\ref{eq:enNum}) and (\ref{eq:enDen}) simplify as
\begin{equation}
E_n(\tau) = {\sum_i x_i c_{\bm{s}^{(i)}} \left( E_L(\bm{r}^{(i)}, \bm{s}^{(i)}) + \sum_{\bm{s}} T_{\bm{s}^{(i)} \bm{s}} \right)}
\end{equation}
and
\begin{equation}
E_d(\tau) = \sum_i x_i c_{\bm{s}^{(i)}}
\end{equation}
Elements of the ground-state eigenvector of $\mathbf{T}$ are used as the coefficients $\lbrace c_{\bm{s}} \rbrace$.

\section{Convergence Analysis of Diffusion Monte Carlo}
At finite time step and finite walker number, the DMC energy averaged over the trajectory is biased, i.e. it does not converge to the exact energy as the trajectory length is increased. This section describes our efforts to estimate and reduce the magnitudes of these biases for the binding energies reported in the main text.

\begin{figure}[h]
\centering
\includegraphics{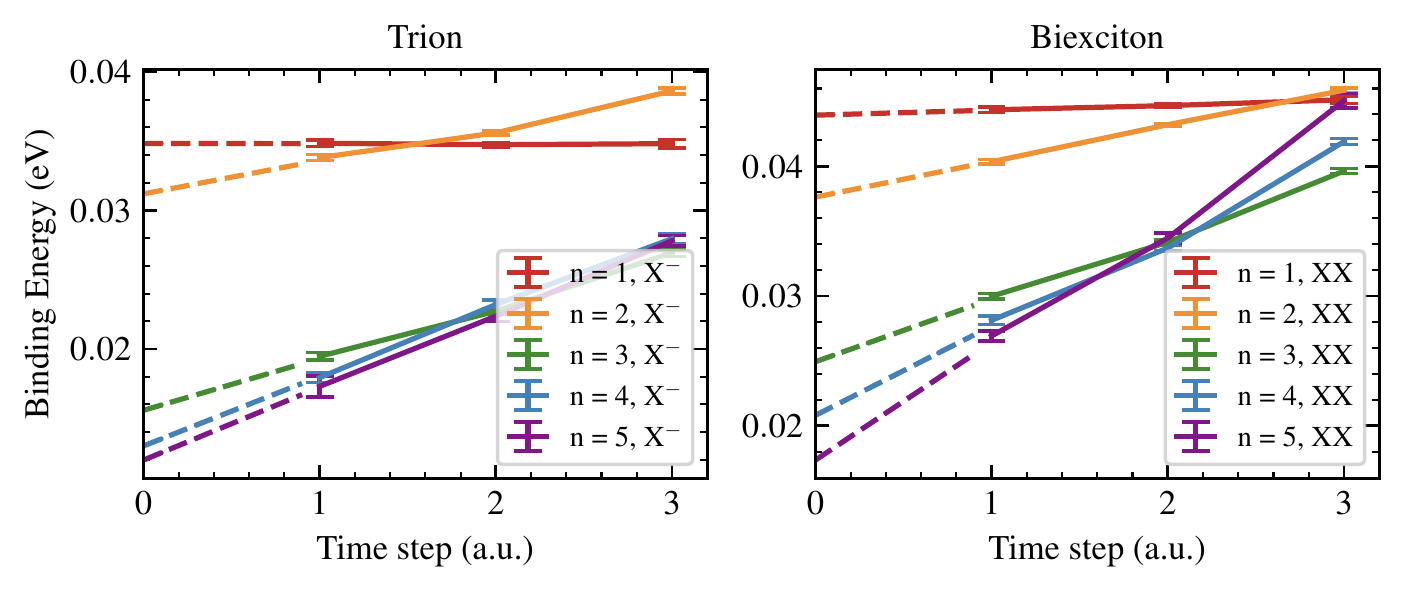}
\caption{Mean binding energies from DMC for trions and biexcitons as a function of time step, at a fixed number of walkers (10,000).}
\label{fig:dmcdt}
\end{figure}

The time step bias arises from both the operator splitting approximation made in (\ref{eq:hopDifEq}) and (\ref{eq:xyDifEq}) and the simplifying approximation made in (\ref{eq:approxHop}). 
The dependence of the mean DMC energy from each trajectory on the time step, $\Delta \tau$, is shown in Figure \ref{fig:dmcdt}. The energy depends more strongly on $\Delta \tau$ for $n>1$, suggesting that the majority of the time step bias arises from the approximation in (\ref{eq:approxHop}). Nevertheless, the dependence is linear in all cases, as in the standard DMC method if sufficiently small time steps are used~\cite{Umrigar1993}.
Linear extrapolation to $\Delta \tau = 0$ was therefore performed to obtain the binding energies reported in the main text.

\begin{figure}[h]
\centering
\includegraphics{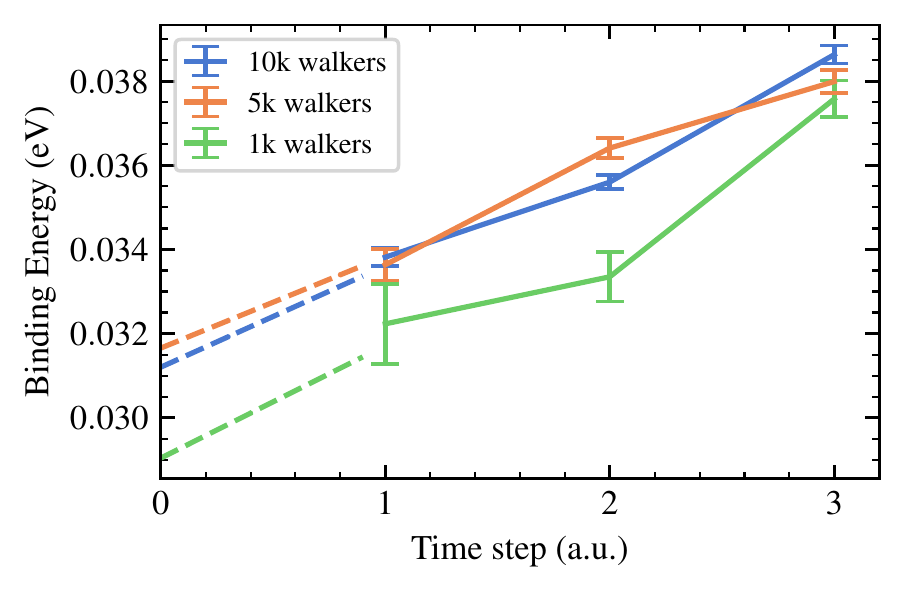}
\caption{Dependence of the trion binding energy for n = 2, calculated at different time steps, on the number of walkers used in DMC calculations.}
\label{fig:dmcwalkers}
\end{figure}

The finite walker bias arises from the procedure for stabilizing the norm of the solution by adjusting the energy shift, as in (\ref{eq:shiftUpdate}). This bias scales as $N_w^{-1}$ in the standard DMC algorithm~\cite{Umrigar1993}.
Mean binding energies for the trion system with $n=2$ with different numbers of walkers are presented in Figure \ref{fig:dmcwalkers}. Mean energies and the extrapolated $(\Delta \tau = 0)$ energy do not change significantly as the number of walkers is increased from 5,000 to 10,000, so 10,000 walkers were used to calculate all results presented in the main text.

\section{Convergence Analysis of Stochatic Variational Method}

\begin{figure}[h]
    \centering
    \includegraphics{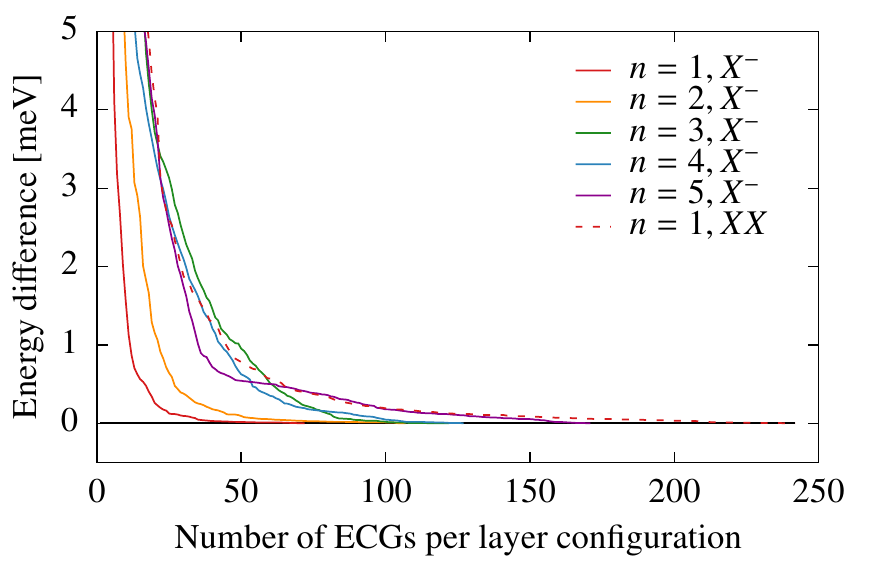}
    \caption{Convergence of the energy calculated with the SVM with repect to the number of ECGs per sublayer configuration.}
    \label{fig:conv}
\end{figure}

In Fig.~\ref{fig:conv}, we show the convergence behavior of the SVM for the trion and biexciton energies.  
As can be seen, the energies
are converged to better than 1~meV with about 50 ECGs per sublayer configuration; for the $n=5$ trion,
this corresponds to approximately $2\times10^3$~ECGs.

\bibliography{bib}